# Scalable synchronization of spin-Hall oscillators in out-of-plane field

V. Puliafito,[1] A. Giordano,[2] A. Laudani,[3] F. Garesci[1], M. Carpentieri,[4] B. Azzerboni,[1,5] and G. Finocchio[2,5 a)]

[1]*Department of Engineering, University of Messina, I-98166 Messina, Italy*

[2]*Department of Mathematical and Computer Sciences, Physical Sciences and Earth Sciences, University of Messina, I-98166 Messina, Italy*

[3]*Department of Engineering, University of Roma Tre, I-00154 Roma, Italy*

[4] *Department of Electrical and Information Engineering, Politecnico di Bari, I-70125 Bari, Italy*

[5]*Istituto Nazionale di Geofisica e Vulcanologia (INGV), Via Vigna Murata 605, 00143 Roma, Italy*

A strategy for a scalable synchronization of an array of spin-Hall oscillators (SHOs) is illustrated. In detail, we present micromagnetic simulations of two and five SHOs realized by means of couples of triangular golden contacts on the top of a Pt/CoFeB/Ta trilayer. Results highlight that the synchronization occurs for the whole current region that gives rise to the excitation of self-oscillations. This is linked to the role of the magnetodipolar coupling, which is the phenomenon driving the synchronization when the distance between oscillators is not too large. Synchronization turns out to be also robust against geometrical differences of the contacts, simulated by considering variable distances between the tips ranging from 100nm to 200nm. Besides, it entails an enlargement of the radiation pattern that can be useful for the generation of spin-waves in magnonics applications. Simulations performed to study the effect of the interfacial Dzyaloshinskii-Moriya interaction show nonreciprocity in spatial propagation of the synchronized spin-wave. The simplicity of the geometry and the robustness of the achieved synchronization make this design of array of SHOs scalable for a larger number of synchronized oscillators.

---

a) Author to whom correspondence should be addressed. Electronic mail: gfinocchio@unime.it.

Microwave oscillators are widely employed in modern technology.[1] For example, in wireless high-speed communications, they provide the clocking of the systems, as well as the generation of the carrier waves. The most common type of semiconductor microwave oscillators is the voltage-controlled oscillator (VCO).[2] It exhibits high operating frequencies (over 100 GHz), low cost, and low power consumption, nonetheless it holds limited tunability (±20%).[3, 4] Microwave spin-transfer-torque nano-oscillators (STNOs)[5, 6, 7, 8] and spin-Hall oscillators (SHOs)[9, 10, 11, 12] seem to be promising as solutions beyond VCOs that are also compatible with complementary-metal-oxide silicon (CMOS) technology. In addition to low cost, low power consumption and high output frequencies, they offer a tunability, wider than VCO, on current and magnetic field.[8, 13] Other features are a better scalability (over 50 times smaller) and high quality factors,[14] they work in a broad range of temperatures and show intrinsic radiation hardness. On the other hand, the main weakness of STNOs and SHOs is the low output power (order of microwatts for the magnetic tunnel junction STNO).[13, 15] This limitation can be overcome by the synchronization of a number of oscillators. With this regard, Kaka et al.[16] and Mancoff et al.[17] authored two milestone papers where the first experimental observation of two phase-locked nano-contact STNOs was demonstrated. The synchronization was considered mutual since each oscillator participated actively in the phenomenon, and, among the results, it entailed the desired increase of the output power and a reduction of linewidth. Starting from those papers, synchronization of STNOs has been diffusely studied, theoretically,[18, 19] numerically,[20, 21] and experimentally.[22, 23, 24] Ruotolo et al.,[23] for example, observed the synchronization of four closely-spaced vortex-based STNOs with no need of external field, whereas Houshang et al.[24] recently demonstrated the synchronization of five STNOs by controlling, through a combination of Oersted field and external in-plane field, a highly directional spin-wave beam.[25] Concerning synchronization of SHOs, Demidov et al.[26] demonstrated the injection locking[27, 28, 29] of an SHO to a microwave current, while synchronization of two nanoconstriction-based[30] SHOs has been predicted by Kendziorczyk et al..[31]

In this Letter, we predict a strategy that permits to achieve a scalable (in terms of number) synchronization of SHOs considering a generalization of the geometry introduced in Ref. 11. Our key result is that synchronization always takes place when the distance between SHOs is not large enough. In other words, the magnetization precession of all the involved spins is frequency locked for the whole range of current starting from the critical excitation value. This behavior is linked to the mechanism driving the synchronization that is the magnetodipolar coupling. Synchronization is also robust against geometrical variations of the oscillators and/or the effect of the interfacial Dzyaloshinskii-Moriya interaction (IDMI).[32] The synchronized mode is characterized by a unique radiation pattern, as large as the number of synchronized oscillators increase. This result is very promising for the realization of spin wave generators in magnonics applications. Although the synchronization of five SHOs is presented in detail here, the number can be increased to a much larger value.

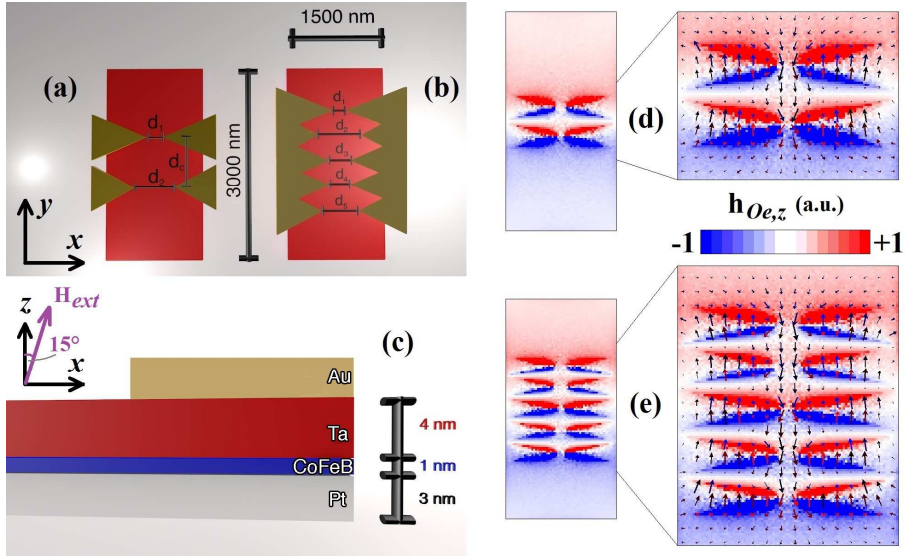

FIG. 1. Sketch of the devices under investigation: (a) two SHOs and (b) five SHOs realized by means of couples of Au triangular contacts over a trilayered stack Pt/CoFeB/Ta (c). Reference systems are also shown. (d)-(e) Oersted field distribution corresponding to: (d) 2 SHOs, $I$=2.56mA, (e) 5 SHOs, $I$=4.72mA.

Figs. 1(a)-(c) display the geometrical details of the two devices under investigation, two and five SHOs are built on top of a Pt(3)/CoFeB(1)/Ta(4) (thicknesses in nm) multilayer with a rectangular cross section of 1500x3000nm$^2$. Differently from the typical heavy-metal/ferromagnet/oxide layered structure of a SHO, here the ferromagnet (CoFeB) is sandwiched between two heavy metals (Pt and Ta). This option arises from the need to reduce the critical current to excite propagating modes in an SHO that are still not observed experimentally. In fact, in this system, the spin-Hall effect (SHE) efficiency is increased. Pt and Ta have an opposite sign of the spin-Hall angle and, as demonstrated by Woo *et al.*[33], the top and bottom interfaces work in concert and enhance the total torque on the magnetization (typical values of spin-Hall angles for Pt and Ta are ~+0.08[34] and ~-0.15,[35, 36] respectively, whereas a total spin-Hall angle of 0.34 was observed in the sandwiched configuration).[33] To realize an SHO in this system, two triangular Au contacts (150nm thick) are deposited on top (similarly to the strategy developed in Ref. 11), in this way the current is locally injected in the center of the ferromagnet[12]. The realization of *N* oscillators can be made through *N* couples of contacts (see Fig. 1(b)). The distance between each couple of contacts, namely between each oscillator, is indicated with $d_c$ and, in this Letter, we show results for $d_c$ = 400 and 800nm. Distances between contacts are indicated with $d_i$ ($i$=1,...,5) and are differentiated in order to model fabrication induced geometrical differences in an array of SHOs. As reference, in the paper by Demidov *et al.*[11] the distance between the contacts

is 100nm, in our study we choose 100nm≤$d_i$≤200nm. In detail, for the device with two contacts, we analyze four different cases: (*i*) $d_1$=100nm, $d_2$=200nm, $d_c$=400nm, (*ii*) $d_1$=100nm, $d_2$=110nm, $d_c$=400nm, (*iii*) $d_1$=100nm, $d_2$=150nm, $d_c$=400nm, (*iv*) $d_1$=100nm, $d_2$=200nm, $d_c$=800nm. On the other hand, the device with five oscillators is characterized by $d_1$=170nm, $d_2$=200nm, $d_3$=100nm, $d_4$=150nm, $d_5$=120nm, $d_c$=400nm. A single oscillator with a similar geometry, but with a ferromagnet/heavy-metal bilayer, has been already studied by our group where it has been identified clearly for which external fields it is possible to have the excitation of propagating spin-waves.[12, 37]

Micromagnetic simulations are performed by means of our self-implemented code[38] that solves the Landau-Lifshitz-Gilbert equation of motion, where the SHE is included as an additional term modeled as Slonczewski-type torque:[5, 37, 39]

$$\frac{d\mathbf{m}}{d\tau} = -\mathbf{m} \times \mathbf{h_{EFF}} + \alpha_G \mathbf{m} \times \frac{d\mathbf{m}}{d\tau} - \frac{g\mu_B}{2\gamma_0 e M_S^2 t_{CoFeB}} \alpha_H \mathbf{m} \times \mathbf{m} \times (\hat{z} \times \mathbf{J}) \qquad (1)$$

where **m** and **h**$_{EFF}$ are the normalized magnetization and effective field of the ferromagnet, respectively. **h**$_{EFF}$ includes the standard magnetic field contributions, as well as the IDMI and the Oersted field (see its spatial distribution in Figs. 1(d)-(e)). $\tau$ is the dimensionless time $\tau = \gamma_0 M_S t$, where $\gamma_0$ is the gyromagnetic ratio, and $M_S$ is the saturation magnetization of the ferromagnet. $\alpha_G$ is the Gilbert damping, $g$ is the Landè factor, $\mu_B$ is the Bohr Magneton, $e$ is the electron charge, $t_{CoFeB}$ is the thickness of the ferromagnetic layer, $\alpha_H$ is the spin-Hall angle obtained from the ratio between the spin current and the electrical current. $\hat{z}$ is the unit vector of the out-of-plane direction and **J** is the spatial distribution of the current density in the heavy metals, computed by averaging the current densities flowing in the Pt and Ta over the two sections respectively. The IDMI is included in the effective field as $\mathbf{h}_{IDMI} = -\frac{2D}{\mu_0 M_S^2}\left[(\nabla \cdot \mathbf{m})\hat{z} - \nabla m_z\right]$, where $m_z$ is the *z*-component of the normalized magnetization and $D$ is the IDMI parameter. The boundary condition for the exchange interaction that takes into account the presence of the IDMI is $d\mathbf{m}/dn = -(1/\chi)(\hat{z} \times \mathbf{n}) \times \mathbf{m}$ where **n** is the unit vector normal to the edge, $\chi = \frac{2A}{D}$ is a characteristic length related to the IDMI and $A$ is the exchange constant.[36, 40] For the simulations discussed in this Letter, we have considered the following physical parameters: $M_S$=1x10$^6$A/m, $A$=2.0x10$^{-11}$J/m[41], interfacial perpendicular anisotropy induced at the boundaries between CoFeB and the heavy metals characterized by the anisotropy constant $K_u$=5.5x10$^5$J/m$^3$,[42] $\alpha_G$ =0.03, and $\alpha_H$ =0.34.[33] For the simulations including IDMI, we have chosen a value of 1.5mJ/m$^2$ for the parameter $D$.[37] An external field of 400mT, tilted 15° degrees with respect to the *z*-axis, is applied to the device and tilts the equilibrium magnetization at about 23° with respect to the *z*-axis. Snapshots of the Oersted field in the ferromagnet are shown in Figs. 1(d)-(e) for the devices with two and five SHOs, respectively. These computations are based on the numerical framework already described in previous works[12, 37] and they have been performed within a parallel processing framework which has been designed and implemented for accelerating algorithms computation.[43, 44]

Fig. 2 summarizes the simulation results of the two SHOs. The output frequencies of the device *vs.* applied current are reported for the abovementioned four cases (*i*) - (*iv*), with and without the inclusion of the IDMI (Figs. 2(a)-(d)). In any of the four cases and for the whole range of applied currents (starting from the threshold of the self-oscillations), we observe that the spins oscillate at the same frequency (frequency locked), with a spatially dependent phase shift, giving rise to a single frequency peak in the power spectrum (see supplementary material Note 1 for the spatially dependent phase shift). Synchronization still takes place for any applied current also in presence of a small variation of out-of-plane amplitude and in-plane field angle (not shown).

Fig. 2(e) shows the results for the case (*iii*) and highlights that the wave vectors are almost the same along the four directions, namely the spin-wave radiation is isotropic (the excitation area is elongated due to the geometry of the contacts). This result, together with the synchronization at any current value, makes the double oscillator comparable to a single oscillator with a wider excitation area (its tunability is about 2-2.5 GHz). Figs. 2(a)-(d) also show that frequency of synchronized mode exhibits blue shift as a function of the applied current as expected for Slonczewski-like spin-waves. With this regard, Figs. 2(f)-(g) show the snapshots of the magnetization configuration for the geometry of case (*iii*) (corresponding to Fig. 2(c)), without and with IDMI (Multimedia wiew Movie 2(f) and 2(g)). In both cases, the radiation pattern is unique. The inclusion of IDMI in the model introduces further nonlinearities in the behavior of the device, and the output frequency shows a non-monotonic behavior (see, in particular, Fig. 2(d)). Concerning the mode profile, it is clearly observed that is non reciprocal, in agreement with previous results on the nonreciprocal spin-wave propagation in presence of IDMI.[45, 46]

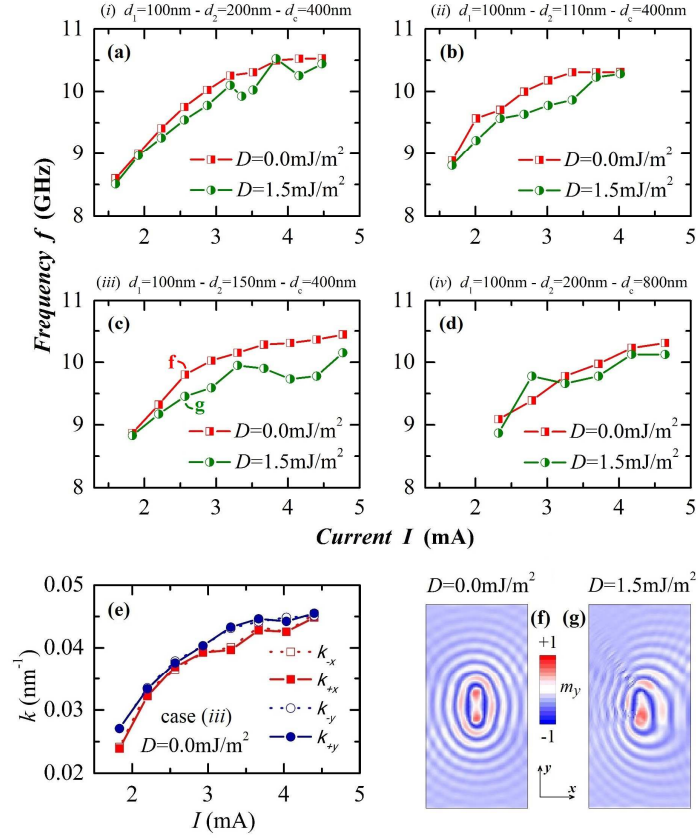

FIG. 2. (a)-(d) Magnetization oscillation frequency of two synchronized SHOs for four different configurations of the Au contacts. In each graph, frequencies with $D=0$mJ/m$^2$ and with $D=1.5$mJ/m$^2$ are reported. (e) Wave vectors along the in-plane axes for the case in (c). (f)-(g) Snapshots of the magnetization corresponding to the points "f" and "g" indicated in (c) for $I=2.56$mA (see Multimedia view Movie 2(f) and 2(g)).

In order to investigate on the phenomenon driving the synchronization, we have also simulated an ideal experiment where the magnetic region between the two SHOs has been cut in order to avoid spin-wave coupling. In those computations, synchronization takes place highlighting the key-role of magnetodipolar coupling as driving phenomenon (see supplementary material Note 2 for the simulations with the cut ferromagnet). This is in agreement with a previous work, where it is shown that synchronization driven by magnetodipolar coupling is characterized by a unique wavefront as in our cases (see Fig. 5 of Ref. 21). The leading role of magnetodipolar coupling also explains the occurrence of synchronization for all the current values, that is linked to the distance $d_c$ between the SHOs. Simulations performed with $d_c=1600$nm, scenario where the synchronization is driven mainly by spin wave interactions, in fact, show locking for a limited range of current (see supplementary material Note 3 for an example of unsynchronized state at $d_c=1600$nm). Synchronization, lastly, is also mutual, since the frequency of the locked mode is in-between the frequencies of the two single SHOs (see supplementary material Note 4 for the evaluation of the frequencies of the single SHOs).

Since the first work on the synchronization of two STNOs,[16] it was necessary more than 10 years to observe the synchronization of five STNOs. That has required larger contacts, a well-defined spatial alignment, and the intuition of the role of the Oersted field.[24] Encouraged by the results of the device with two SHOs, we have performed micromagnetic simulations increasing the number $N$ of SHOs up to $N=5$. Our computations show synchronization characterized by similar properties of the 2 SHOs. In particular, Fig. 3(a) shows the frequency of the synchronized mode for $N=5$, with and without IDMI. Also in this configuration, we observe the frequency blue shifts with the applied current, and oscillators are synchronized for the whole range of currents (mind that $d_c=400$nm). Once again, IDMI distorts the profile of the spin-wave, which, even if it remains unique and continuous, becomes non-reciprocal (see snapshots in Figs. 3(b)-(c), and the corresponding Multimedia view Movie 3(b) and 3(c)), for the dynamics without and with IDMI, respectively). On the whole, the excitation area is larger than the case of two oscillators, due to the geometry of the contacts. Differently from Ref. 24, in the framework studied in this Letter the Oersted field is symmetric and has a negligible role. This feature promises a propagation of the spin-wave at longer distances, that can be useful if spin-wave has to reach other devices on the same ferromagnet, namely for magnonics applications.

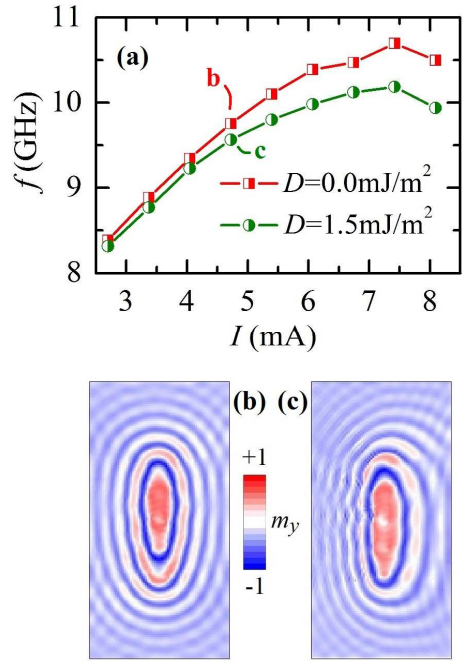

FIG. 3. (a) Frequency of the magnetization oscillation of 5 synchronized SHOs without and with IDMI. Distances between contacts are $d_1=170$nm, $d_2=200$nm, $d_3=100$nm, $d_4=150$nm, $d_5=120$nm, $d_c=400$nm. (b), (c) Magnetization configurations obtained without and with IDMI, respectively, corresponding to the points "b" and "c" indicated in (a) for $I=4.72$mA (see Multimedia view Movie 3(b) and 3(c)).

In conclusion, our numerical simulations have demonstrated the synchronization of two and five SHOs. In our geometry, synchronization occurs for the whole range of current if the distance between SHOs is not too large, and, in our case, this critical distance is between 800 and 1600nm. The main mechanism at the basis of the synchronization is a strong coupling due to the large and long range magnetodipolar fields that in our scenario are largest as compared to previous works on synchronization. This gives rise to a robust synchronization against differences in the geometry of the contacts. In the synchronized state, an enlargement of the isotropic/anisotropic radiation pattern in the absence/presence of IDMI, as a function of the increase of the number of locked oscillators, can be useful for the realization of spin-wave source for magnonic applications. The robust occurrence of the synchronization makes our design of SHOs array scalable for a larger number of oscillators.

See supplementary material for details on the spatially dependent phase shift, the simulations with the cut ferromagnet, an example of unsynchronized state at $d_c$=1600nm, and the evaluation of the frequencies of the single SHOs.

This work was supported by the project PRIN2010ECA8P3 from Italian MIUR and the executive programme of scientific and technological cooperation between Italy and China for the years 2016-2018 (code CN16GR09) funded by Ministero degli Affari Esteri e della Cooperazione Internazionale. The authors thank Domenico Romolo for the graphical support.

# SUPPLEMENTARY NOTES

# Scalable synchronization of spin-Hall oscillators in out-of-plane field

V. Puliafito, A. Giordano, A. Laudani, F. Garescì, M. Carpentieri, B. Azzerboni, and G. Finocchio

## Supplementary Note 1

**The spatially dependent phase shift in the synchronized mode.** Fig. S1(a) shows the spatial distribution of the current density (*x*-component) flowing in the Pt layer for the case of two SHOs at a distance $d_c$=800nm (case (*iv*) of the main text). In the panel are also indicated 5 points (A-E) linked to the CoFeB layer, points at which we have computed the time behavior of the *y*-component of the magnetization (Fig. S1(b)). It is clear that those 5 curves are synchronized with a spatially dependent phase shift. In particular the phase shift is small if we consider cells close to only one spin-Hall oscillator (SHO) (A-B and D-E), whereas it is large if we consider the cell C between the SHOs (B-C and C-D).

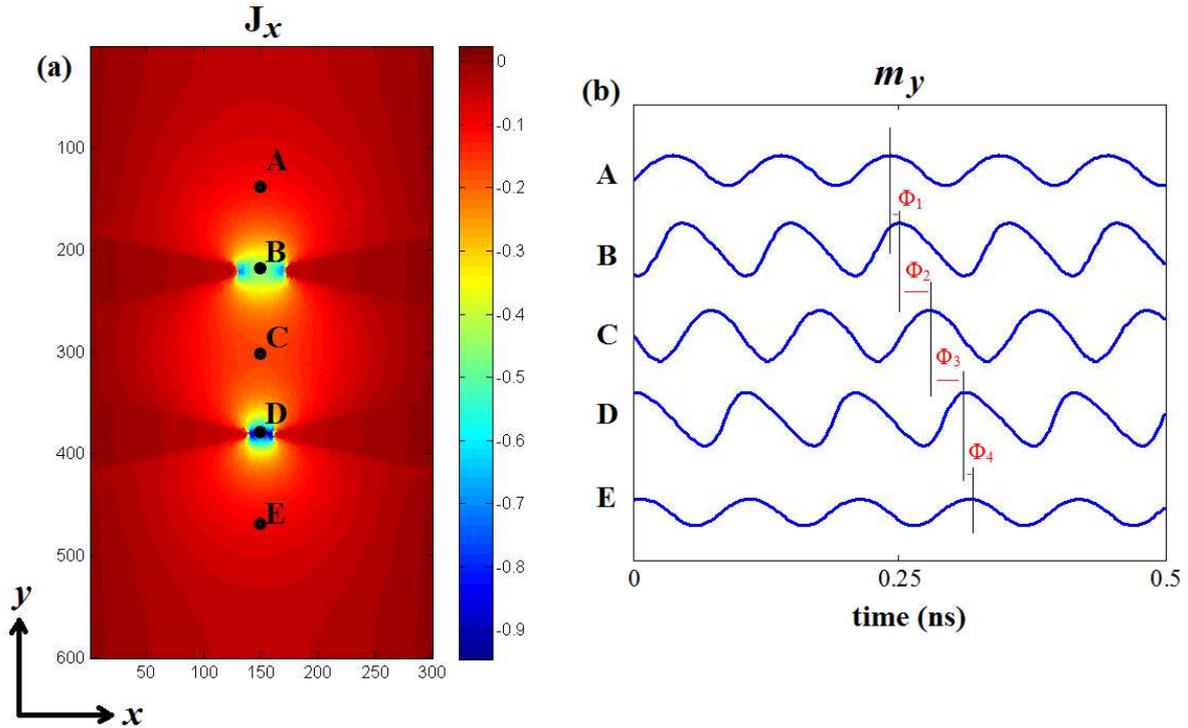

FIG. S1. (a) Colorplot of the *x*-component of the current density *J* flowing in the Pt layer for the case of two SHOs at a distance $d_c$=800nm. A, B, C, D, and E indicate 5 different cells of our computational area related to the CoFeB. (b) Time behavior of the *y*-component of the magnetization in the five cells of (a). The phase shifts are indicated to point out that curves are synchronized with a spatially dependent phase shift.

## Supplementary Note 2

**The leading role of magnetodipolar coupling in the synchronization.** Synchronization of oscillators can be driven by different physical phenomena, already discussed in several papers, such as spin-wave interaction, magnetodipolar coupling, spin-wave beams. One of the most efficient techniques to reveal the importance of those phenomena in the occurrence of synchronization is the physical separation of the oscillators, that can be realized by means of a cut in the ferromagnet between the oscillators. In that circumstance, spin-wave interaction is prevented. We have performed such an ideal experiment by means of our simulations (see Fig. S2). In those computations, synchronization still occurs highlighting that the magnetodipolar coupling is the key factor to achieve the synchronization in our array of SHOs for $d_c$=400nm and $d_c$=800nm (see supplementary material Movie S1). These results are also in agreement with previous works about STOs (see for example Ref. 21 of the main text, in particular Fig. 5).

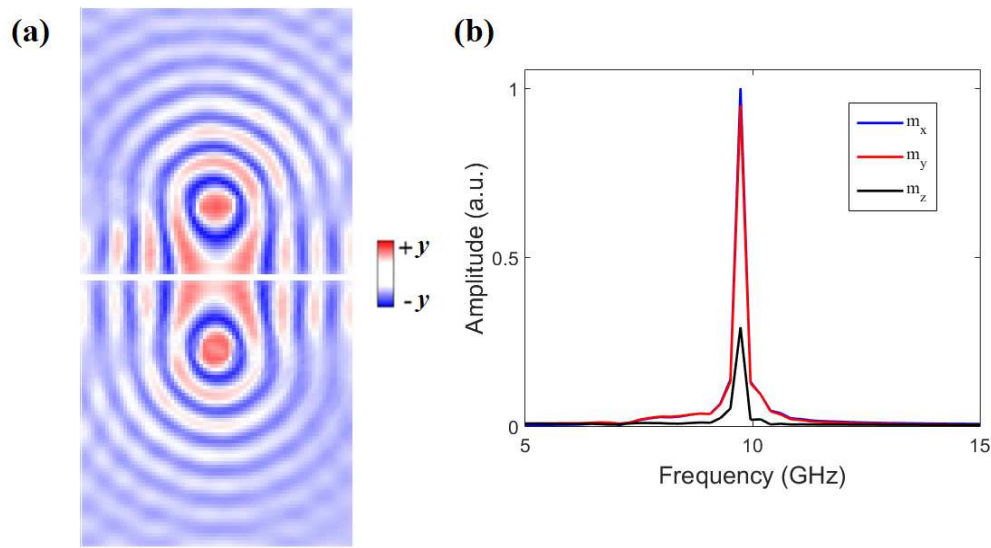

FIG. S2. Synchronization of two separated SHOs driven by magnetodipolar coupling. (a) An example of magnetization configuration of the mode resulting from the synchronization of two SHOs at a distance of 800nm (case (*iv*) of the main text, *I*=3.25mA). Color refers to the *y*-component of the magnetization, blue negative, red positive. Ferromagnet is cut to avoid spin-wave interaction. Nonetheless a unique wavefront is formed, and (b) a unique peak is detected in the frequency spectrum. See also supplementary material Movie S1.

## Supplementary Note 3

**The configuration of two SHOs at a distance $d_c$=1600nm.** Simulations discussed in the main text show synchronization of two and five SHOs for any value of current, starting from the threshold current for the oscillations. In those simulations, the distance between the oscillators ($d_c$) is 400 or 800nm. It is expected that for some larger value of $d_c$, synchronization occurs either for a limited range of currents or for no values of current. In order to investigate on this aspect, we have performed some simulations of two SHOs at a distance $d_c$=1600nm. Those additional simulations have pointed out that the synchronization is not achieved for all the current values and the results are similar to what observed in spin-torque oscillators (STOs) (Refs. 17-18 of the main text). In particular, we have identified a key difference: for SHOs at a distance of 400 and 800nm the radiation pattern is unique (Fig. S3 (a)) while for SHOs at a distance of 1600nm two radiation patterns can be clearly observed (Fig. S3(b)). In conclusion, the SHOs are not synchronized at all the currents (see Fig. S3(c) for an example of Fourier transform of unlocked SHOs achieved at $d_c$=1600nm and $I$=3.88mA) when the distance between them is large enough. In our case this critical distance is between 800 and 1600nm.

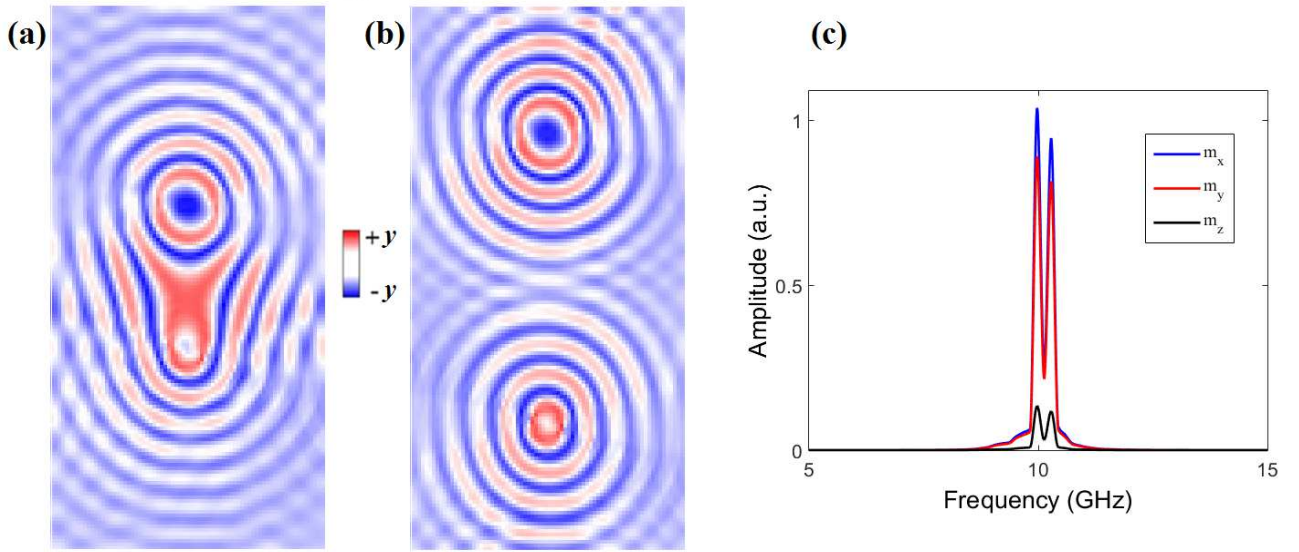

FIG. S3. Magnetization configuration for two SHOs with (a) $d_c$=800nm (case (*iv*) of the main text) and (b) $d_c$=1600nm. Color refers to the y-component of the magnetization, blue negative, red positive. In the first case a unique radiation pattern is created as a consequence of synchronization. In the second case, two different radiation patterns can be clearly observed and they synchronize in a limited range of current. (c) Fourier transform of the magnetization dynamics for 2 SHOs for $d_c$=1600nm and $J$=1.0x10$^9$ A/cm$^2$, corresponding to $I$=3.88mA.

## Supplementary Note 4

**Frequencies of the single SHOs and of the synchronized mode.** Fig. S4(a) shows the frequencies of two single SHOs, with the distance between the tips of 100 and 200nm, as a function of the current density $J$. It is possible to observe that the frequency decreases with the decrease of that distance. In the same figure, we have included the frequency of the synchronized mode, as a function of $J$, for the case (*iv*) of the main text ($d_1$=100nm, $d_2$=200nm, $d_c$=800nm). That frequency is unexpectedly lower than the other two. Nonetheless, it is important to underline that current density is not uniform in the device and the value of $J$ indicated in the graphs is simply the maximum amplitude of current density in the device. In order to better evaluate the frequency of the single SHOs, we have performed simulations using the current density distribution for the case (*iv*) but removing for a set of simulations the upper part of ferromagnet and for another set of simulations the lower one, so that the SHOs with tips at distance 100nm and 200nm can be studied singularly (mind that we left the location of the tips exactly at the same place of the case (*iv*) thus are not at the center of the new domain). In Fig. S4(b) we have used filled symbols to indicate the corresponding frequencies of the two single oscillators for two different values of $J$ (see Figs. S4(c) and (d) for the magnetization configurations of the two single SHOs, and supplementary material Movie S2). If we look at these frequencies, the frequency of the synchronized mode, reported here again as in (a), is in the middle of them, as it typically occurs for the synchronization of different oscillations.

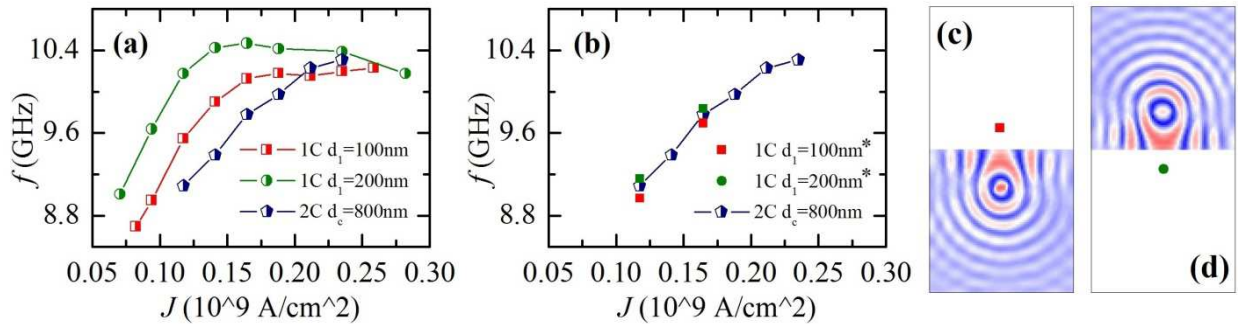

FIG. S4. (a) Frequency of 2 single SHOs with tips at distance 100 and 200nm, as a function of the current density $J$, compared with the frequency of the synchronized mode for the case (*iv*) of the main text (2 SHOs at a distance $d_c$=800nm). (b) Filled symbols to represent the frequencies of the single SHOs calculated with the same current density distribution of the case (*iv*) and by removing half of the ferromagnet, as shown in (c) and (d), compared with the frequency of the synchronized mode as in (a). See supplementary material Movie S2 for the dynamics corresponding to (c).